\documentclass[useAMS,usegraphicx,usenatbib]{mn2e}          

\usepackage{times}

\newcommand{\Msun}{\ensuremath{\,{\rm M}_\odot}}                  
\newcommand{\Rsun}{\ensuremath{\,{\rm R}_\odot}}                  
\newcommand{\Teff}{\ensuremath{T_{\rm eff}}}                      
\newcommand{\logg}{\ensuremath{\log g}}                           
\newcommand{\Mjup}{\ensuremath{\,{\rm M}_{\rm Jup}}}              
\newcommand{\Rjup}{\ensuremath{\,{\rm R}_{\rm Jup}}}              
\newcommand{\Teq}{\ensuremath{T_{\rm eq}^{\,\prime}}}             
\newcommand{\safronov}{\ensuremath{\Theta}}                       
\newcommand{\kms}{\,km\,s$^{-1}$}                                 
\newcommand{\ms}{\,m\,s$^{-1}$}                                   
\newcommand{\mss}{\,m\,s$^{-2}$}                                  
\newcommand{\as}{\ensuremath{^{\prime\prime}}}                    
\newcommand{\FeH}{\ensuremath{\left[\frac{\rm Fe}{\rm H}\right]}} 
\newcommand{\pjup}{\ensuremath{\,\rho_{\rm Jup}}}                 
\newcommand{\psun}{\ensuremath{\,\rho_\odot}}                     
\newcommand{\chir}{\ensuremath{\chi_\nu^{\,2}}}                   
\newcommand{\mc}[1]{\multicolumn{2}{c}{#1}}
\newcommand{\mcc}[1]{\multicolumn{3}{c}{#1}}

\newcommand{\erc}[3]{\mc{\ensuremath{#1^{+#2}_{-#3}}}}

\newcommand{\ermcc}[5]{\mcc{\ensuremath{{#1\,^{+#2}_{-#3}}\,^{+#4}_{-#5}}}}
\newcommand{\dms}[3]{\ensuremath{#1^\circ\,#2'\,#3''}}
\newcommand{\reff}[1]{{#1}}                                   

\title[The physical properties of HAT-P-5]
      {Physical properties and radius variations in the HAT-P-5 planetary system from simultaneous four-colour photometry}

\author[Southworth et al.]
       {John Southworth\,$^{1}$, L.\ Mancini\,$^{2}$, P.\ F.\ L.\ Maxted\,$^{1}$, I.\ Bruni\,$^{3}$, J.\ Tregloan-Reed\,$^{1}$,
        \newauthor M.\ Barbieri\,$^{4}$, N.\ Ruocco\,$^{5,6}$, P.\ J.\ Wheatley$^{7}$ \\
        $^{1}$\,Astrophysics Group, Keele University, Newcastle-under-Lyme, ST5 5BG, UK \\
        $^{2}$\,Max Planck Institute for Astronomy, K\"onigstuhl 17, 69117 -- Heidelberg, Germany \\
        $^{3}$\,INAF -- Osservatorio Astronomico di Bologna, Via Ranzani 1, 40127 Bologna, Italy \\
        $^{4}$\,Observatoire de la C\^ote d'Azur, 06304 Nice Cedex 4, France \\
        $^{5}$\,Astrocampania, Sez.\ Stabia – Penisola Sorrentina, Italy \\
        $^{6}$\,Unione Astrofili Italiani, Variable Stars Division, Italy \\
        $^{7}$\,Department of Physics, University of Warwick, Coventry, CV4 7AL, UK
}

\begin{document} \maketitle 

\begin{abstract}
The radii of giant planets, as measured from transit observations, may vary with wavelength due to Rayleigh scattering or variations in opacity. Such an effect is predicted to be large enough to detect using ground-based observations at multiple wavelengths. We present defocussed photometry of a transit in the HAT-P-5 system, obtained simultaneously through Str\"omgren $u$, Gunn $g$ and $r$, and Johnson $I$ filters. Two more transit events were observed through a Gunn $r$ filter. We detect a substantially larger planetary radius in $u$, but the effect is greater than predicted using theoretical model atmospheres of gaseous planets. This phenomenon is most likely to be due to systematic errors present in the $u$-band photometry, stemming from variations in the transparency of Earth's atmosphere at these short wavelengths. We use our data to calculate an improved orbital ephemeris and to refine the measured physical properties of the system. The planet HAT-P-5\,b has a mass of $1.06 \pm 0.11 \pm 0.01$\Mjup\ and a radius of $1.252 \pm 0.042 \pm 0.008$\Rjup\ (statistical and systematic errors respectively), making it slightly larger than expected according to standard models of coreless gas-giant planets. Its equilibrium temperature of $1517 \pm 29$\,K is within 60\,K of that of the extensively-studied planet HD\,209458\,b.
\end{abstract}



\section{Introduction}                                                                                                              \label{sec:intro}

\begin{table*} \centering
\caption{\label{tab:obslog} Log of the observations presented in this work. $N_{\rm obs}$ is the number
of observations and `Moon illum.' is the fractional illumination of the Moon at the midpoint of the transit.
The aperture sizes are the radii of the software apertures for the star, inner sky and outer sky, respectively.}
\begin{tabular}{llccccccccc} \hline
Transit & Date & Start time & End time & $N_{\rm obs}$ & Exposure & Filter &      Airmass      & Moon   & Aperture   & Scatter \\
        &      &    (UT)    &   (UT)   &               & time (s) &        &                   & illum. & sizes (px) & (mmag)  \\
\hline
1 & 2010 08 23 & 20:24 & 01:33 & 201 &    60    & Str\"omgren $u$ & 1.00 $\to$ 2.01            & 0.994 & 17, 30, 50 & 3.61 \\
1 & 2010 08 23 & 20:24 & 01:33 & 204 &    60    & Gunn $g$        & 1.00 $\to$ 2.01            & 0.994 & 30, 45, 70 & 1.20 \\
1 & 2010 08 23 & 20:24 & 01:33 & 204 &    60    & Gunn $r$        & 1.00 $\to$ 2.01            & 0.994 & 30, 45, 70 & 0.93 \\
1 & 2010 08 23 & 20:24 & 01:33 & 199 &    60    & Johnson $I$     & 1.00 $\to$ 2.01            & 0.994 & 27, 45, 65 & 1.27 \\
2 & 2011 05 26 & 21:18 & 02:47 & 135 & 120--100 & Gunn $r$        & 1.38 $\to$ 1.01 $\to$ 1.07 & 0.294 & 16, 26, 45 & 0.75 \\
3 & 2011 07 19 & 20:39 & 02:04 & 176 &  90--33  & Gunn $r$        & 1.03 $\to$ 1.01 $\to$ 1.53 & 0.861 & 20, 30, 50 & 1.53 \\
\hline \end{tabular} \end{table*}

Of the known extrasolar planets, those which transit their parent stars comprise a rich source of information on the formation and evolution of gas giants. The physical properties of transiting extrasolar planets (TEPs) can be determined -- with a little help from stellar evolutionary theory -- to precisions approaching a few percent\footnote{See: {\tt http://www.astro.keele.ac.uk/$\sim$jkt/tepcat/}} \citep[e.g.][]{Torres++08apj,Me09mn}. The radii of the known TEPs are of particular interest as indicators of the internal structure and heating mechanisms of irradiated giant planets \citep[e.g.][]{Fortney++07apj,Batygin++09apj}. Also, the opacities of different species in the atmospheres of TEPs change with wavelength, causing variations of their radii as measured from transit light curves. These wavelength-dependent variations can be used to probe the atmospheric composition of the objects \citep[see][]{Burrows+07apj}.

\citet{Fortney+08apj} divided the known gas-giant TEPs into two classes according to the presence of appreciable atmospheric opacity from oxidised titanium and vanadium. They calculated the observed radius of a planet with a radius of 1.20\Rjup\ at 1\,bar pressure and a surface gravity of 15\mss, for wavelengths of 0.3--30\,$\mu$m. pM-class planets have high TiO and VO opacity leading to temperature inversions and an increased transit-derived radius over a broad range of optical wavelengths. These objects are predicted to have apparent radii which are the smallest at the bluest optical wavelengths ($<$370\,nm) and the greatest at 450-700\,nm, the difference being 4\% for the 1.20\Rjup\ planet and even larger for one with a lower surface gravity (larger atmospheric scale height). By contrast, the pL planets are less highly irradiated and have negligible TiO and VO opacity. Their transit radii are predicted to be smaller and less variable, with narrow peaks around the Na\,I 589\,nm and K\,I 768\,nm doublets. A similar variation for pL planets was found by \citet{Burrows+08apj} for the case of HD\,209458.

\citet{Zahnle+09apj} has shown that sulphur is an important element for opacity in exoplanet atmospheres, and might serve as a probe of the metallicity of TEPs. S$_2$ opacity becomes large at 240--340\,nm for temperatures above 1200\,K, and HS absorbs strongly in the wavelength interval 300--460\,nm. These effects should lead to large apparent radii at blue-optical and ultraviolet wavelengths. Another relevant phenomenon is Rayleigh scattering, which is proportional to $\lambda^{-4}$ and so should cause the transit radii of TEPs to be larger at shorter wavelengths. Both effects run opposite to the predictions of \citet{Fortney+08apj}; it is not yet clear which processes dominate the measured radii of exoplanets.

The two well-studied TEPs, in terms of variation of radius over optical wavelengths, are HD\,209458\,b and HD\,189733\,b. For the former, \citet{Knutson+07apj} presented HST grism photometry of two transits divided into ten wavelength intervals covering 290--1030\,nm. Differences of radius with wavelength are unclear from these data: \citet{Barman07apj} and \citet{Sing+08apj} found significant variations whereas \citet{Knutson+07apj} and \citet{Me08mn} did not. \citet{Charbonneau+02apj} has found evidence for Na\,I absorption in the planetary atmosphere based on an earlier dataset \citep{Brown+01apj}. These characteristics point to a pL classification for HD\,209458\,b; its moderate equilibrium temperature of $\Teq = 1459 \pm 12$\,K \citep{Me10mn} puts it roughly at the boundary between the pL and pM classes. For HD\,189733\,b, HST observations covering 550--1050\,nm \citep{Pont+08mn} and 290--570\,nm \citep{Sing+11mn} returned only gradual radius variations with wavelength, which have been interpreted as indicative of Rayleigh scattering from a high-altitude atmospheric haze. HD\,189733\,b has $\Teq = 1191 \pm 20$\,K \citep{Me10mn} so is firmly in the pL planetary class. There has therefore been no clear detection of TEP radius variations indicative of a pM classification.

HAT-P-5 was discovered by \citet{Bakos+07apj2} and is a comparatively normal TEP system consisting of a 1.2\Msun\ star orbited by a 1.1\Mjup\ planet. Its planet is slightly hotter ($\Teq = 1516 \pm 29$\,K) than HD\,209458\,b so is on the border of the pL and pM classes. \reff{\citet{Krivov+11mn} found a flux excess at 12 and 22\,$\mu$m indicative of the presence of dust rings at a distance fo several AU from the parent star.} In this work we present the first follow-up transit photometry of HAT-P-5 obtained since the discovery paper, covering four optical passbands. We probe for radius variations between these passbands and improve the measurements of the physical properties of the system.


\section{Observations and data reduction}                                                                                             \label{sec:obs}

\begin{figure*} \includegraphics[width=\textwidth,angle=0]{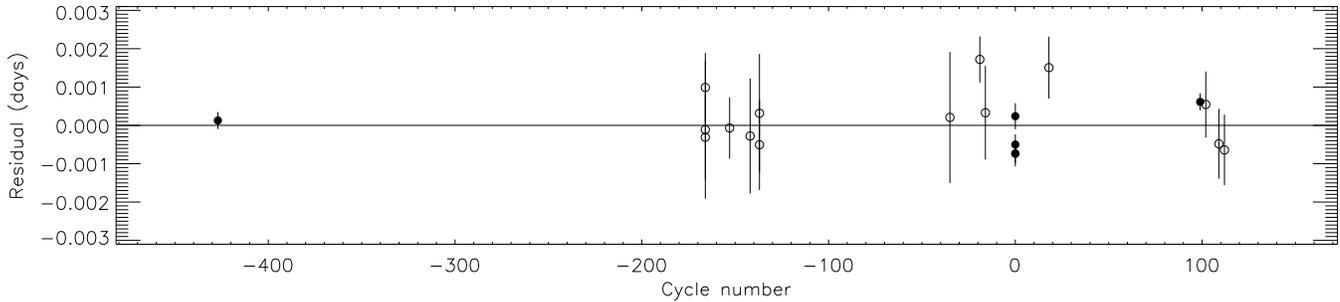}
\caption{\label{fig:minima} Plot of the residuals of the timings of
mid-transit of HAT-P-5 versus a linear ephemeris. The timings obtained
by amateur astronomers are plotted using open circles.} \end{figure*}

\begin{table} \centering \caption{\label{tab:lc} Excerpts of the light
curve of HAT-P-5. The full dataset will be made available at the CDS.}
\setlength{\tabcolsep}{4pt}
\begin{tabular}{llcrr} \hline
Telescope & Filter & BJD(TDB) & Diff.\ mag. & Uncertainty \\
\hline
CAHA22  & $u$ & 2455432.35006 & $-0.00004$ & $0.00316$ \\
CAHA22  & $u$ & 2455432.56365 & $-0.00208$ & $0.00582$ \\[2pt]
CAHA22  & $g$ & 2455432.35006 & $ 0.00156$ & $0.00119$ \\
CAHA22  & $g$ & 2455432.56365 & $ 0.00004$ & $0.00135$ \\[2pt]
CAHA22  & $r$ & 2455432.35006 & $-0.00013$ & $0.00094$ \\
CAHA22  & $r$ & 2455432.56465 & $ 0.00144$ & $0.00099$ \\[2pt]
CAHA22  & $I$ & 2455432.35006 & $ 0.00015$ & $0.00129$ \\
CAHA22  & $I$ & 2455432.56465 & $ 0.00093$ & $0.00134$ \\[2pt]
Cassini & $r$ & 2455708.38967 & $ 0.00066$ & $0.00081$ \\
Cassini & $r$ & 2455708.60776 & $-0.00120$ & $0.00119$ \\[2pt]
Cassini & $r$ & 2455761.36079 & $-0.00104$ & $0.00115$ \\
Cassini & $r$ & 2455761.53645 & $ 0.01876$ & $0.00195$ \\
\hline \end{tabular} \end{table}

We observed one full transit of HAT-P-5 on the night of 2010/08/23, using the 2.2\,m telescope and BUSCA imager at Calar Alto Astronomical Observatory. BUSCA uses dichroics to split the incoming light into four wavelength intervals, which traverse different arms of the instrument and are incident onto four CCDs. This allows photometry to be obtained in four passbands simultaneously. The arms of the instrument cannot be operated individually: their focus and CCD settings are controlled together. The full field of view is 12$\times$12 arcminutes, which the CCDs sample with a plate scale of 0.176\as\ per pixel.

For our observations we chose to have the dedicated BUSCA 109-mm diameter Str\"omgren $u$ and Johnson $I$ filters, in the bluest and reddest arms respectively. For the intermediate arms we opted for 50-mm diameter Gunn $g$ and $r$ filters from the standard Calar Alto filter catalogue. The latter two filters have a better throughput but lead to a reduced field of view: a circle approximately 6 arcmin in diameter. The $g$ and $r$ filters are also thinner than the $u$ and $I$ filters, leading to a difference in focus between the two filter sets. We defocussed BUSCA in such a way as to have the $g$ and $r$ observations more defocussed than those in $u$ and $I$, in order to compensate for the lower throughput of the latter two filters.

The amount of defocussing was set by the requirement that the brightest pixel within the point spread function (PSF) of the target and comparison stars (on all four CCDs) have not more than 35\,000 counts in one observation. An initial exposure time of 90\,s was chosen, which is slightly smaller than our usual value of 120\,s \citep[see][]{Me+09mn,Me+09mn2} to limit the impact of the bright moon. The CCDs were binned 2$\times$2 in order to shorten the readout time. The autoguider was not defocussed along with BUSCA, so could be used to maintain the pointing of the telescope.

We observed two transits of HAT-P-5 in May and July 2011, using BFOSC mounted on the 1.52\,m G.\ D.\ Cassini Telescope\footnote{Information on the 1.52\,m Cassini Telescope and BFOSC can be found at {\tt http://www.bo.astro.it/loiano/}} at Loiano Observatory, Italy. For more information on the use of this instrument for defocussed photometry see \citet{Me+10mn,Me+12b}. Both transits were observed through a Gunn $r$ filter. The second transit was curtailed by cloud before egress. It also suffered from the gradual drift of the telescope back into focus over the night, requiring the exposure time to be successively reduced from the original 90\,s down to 33\,s. As with BUSCA, the telescope pointing was maintained using the autoguider. A summary of the observational data is given in Table\,\ref{tab:obslog}.

One further transit was observed by NR using a 25\,cm aperture Meade LX200 telescope sited near Sorrento, Italy (latitude \dms{40}{37}{07.37} North, longitude \dms{14}{21}{27.46} East, altitude 275\,m above sea level). The CCD used was an SBIG ST7 operating at a plate scale of 1.16\as\,px$^{-1}$, mounted behind an $r$ filter. Data reduction was performed using MaxIm\,DL Pro 5\footnote{\tt http://www.cyanogen.com/} following standard procedures.

Several images were taken with the telescopes properly focussed, to verify that there were no faint stars within the defocussed PSF of HAT-P-5. We find no evidence of a second star near the PSF of our star of interest, and conclude that the transit is not diluted by contaminating light. This agrees with \citet{Daemgen+09aa}, who obtained high-resolution observations of HAT-P-5 as part of a high-speed `lucky' imaging survey of twelve TEP host stars.

Data reduction was undertaken using standard methods. Aperture photometry was performed using the {\sc idl}\footnote{The acronym {\sc idl} stands for Interactive Data Language and is a trademark of ITT Visual Information Solutions. For further details see {\tt http://www.ittvis.com/ProductServices/IDL.aspx}.}/{\sc astrolib}\footnote{\tt http://idlastro.gsfc.nasa.gov/} implementation of {\sc daophot} \citep{Stetson87pasp}. The apertures were placed by hand and shifted to account for pointing variations, which were measured by cross-correlating each image against a reference image. We tried a wide range of aperture sizes and retained those which gave photometry with the lowest scatter compared to a fitted model (see below). We find that the shape of the light curve is very insensitive to the choice of aperture sizes, and to whether flat fields are used in the reduction process. The times of observation were converted from UTC to barycentric Julian date on the TDB timescale, using the {\sc idl} procedures of \citet{Eastman++10pasp}.

Differential photometry was obtained using an optimal ensemble of comparison stars \citep{Me+09mn} combined with a polynomial fit to the data outside transit. The choice of comparison stars has a negligible impact on the shape of the observed transits. A second-order polynomial was required for the CAHA light curves, but a first-order one was enough for the Cassini data. The resulting photometry is given in Table\,\ref{tab:lc}, and reaches a scatter as low as 0.75\,mmag for the first Cassini transit. We also obtained the $z$-band and $R$-band data presented in \citet{Bakos+07apj2} for inclusion in our analysis. These datasets cover two and four transits, respectively.


\section{Analysis}                                                                                                               \label{sec:analysis}

We have measured the physical properties of the HAT-P-5 system using the methods described by \citet{Me08mn,Me09mn,Me10mn,Me11mn}, thus retaining consistency with the {\it Homogeneous Studies} project results. A detailed description of our approach is given by those works, so we only summarise it briefly below.

\subsection{Period determination}                                                                                                 \label{sec:lc:porb}

\begin{table} \begin{center}
\caption{\label{tab:minima} Times of minimum light of HAT-P-5
and their residuals versus the ephemeris derived in this work.}
\setlength{\tabcolsep}{4pt}
\begin{tabular}{l@{\,$\pm$\,}l r r l} \hline
\multicolumn{2}{l}{Time of minimum}      & Cycle  & Residual & Reference \\
\multicolumn{2}{l}{BJD(TDB) $-$ 2400000} & no.    & (JD)     &           \\
\hline
54241.77700 & 0.00022 & -427.0 &  0.00013 & \citet{Bakos+07apj2}           \\   
54969.56817 & 0.00110 & -166.0 & -0.00031 & Gregorio (AXA)                 \\   
54969.56837 & 0.00180 & -166.0 & -0.00011 & Mendez (AXA)                   \\   
54969.56947 & 0.00090 & -166.0 &  0.00099 & Naves (AXA)                    \\   
55005.81857 & 0.00080 & -153.0 & -0.00007 & Norby (AXA)                    \\   
55036.49157 & 0.00150 & -142.0 & -0.00028 & Srdoc (AXA)                    \\   
55050.43371 & 0.00118 & -137.0 & -0.00051 & Br\'at (TRESCA)                \\   
55050.43453 & 0.00155 & -137.0 &  0.00031 & Trnka (TRESCA)                 \\   
55334.85873 & 0.00171 &  -35.0 &  0.00021 & Shadick (TRESCA)               \\   
55379.47582 & 0.00060 &  -19.0 &  0.00172 & Vila'agi (TRESCA)              \\   
55387.83985 & 0.00122 &  -16.0 &  0.00033 & Garlitz (TRESCA)               \\   
55432.45534 & 0.00066 &    0.0 &  0.00024 & This work ($u$)                \\   
55432.45436 & 0.00033 &    0.0 & -0.00074 & This work ($g$)                \\   
55432.45460 & 0.00027 &    0.0 & -0.00050 & This work ($r$)                \\   
55432.45437 & 0.00025 &    0.0 & -0.00073 & This work ($I$)                \\   
55708.51460 & 0.00022 &   99.0 &  0.00061 & This work ($r$)                \\   
55482.64913 & 0.00081 &   18.0 &  0.00151 & Shadic (TRESCA)                \\   
55716.87995 & 0.00086 &  102.0 &  0.00054 & Garlitz (TRESCA)               \\   
55736.39824 & 0.00091 &  109.0 & -0.00048 & Ayiomamitis (TRESCA)           \\   
55744.76350 & 0.00092 &  112.0 & -0.00064 & Garlitz (TRESCA)               \\   
55761.50126 & 0.00290 &  118.0 &  0.00628 & This work (NR)                 \\   
\hline \end{tabular} \end{center} \end{table}

\begin{figure} \includegraphics[width=0.48\textwidth,angle=0]{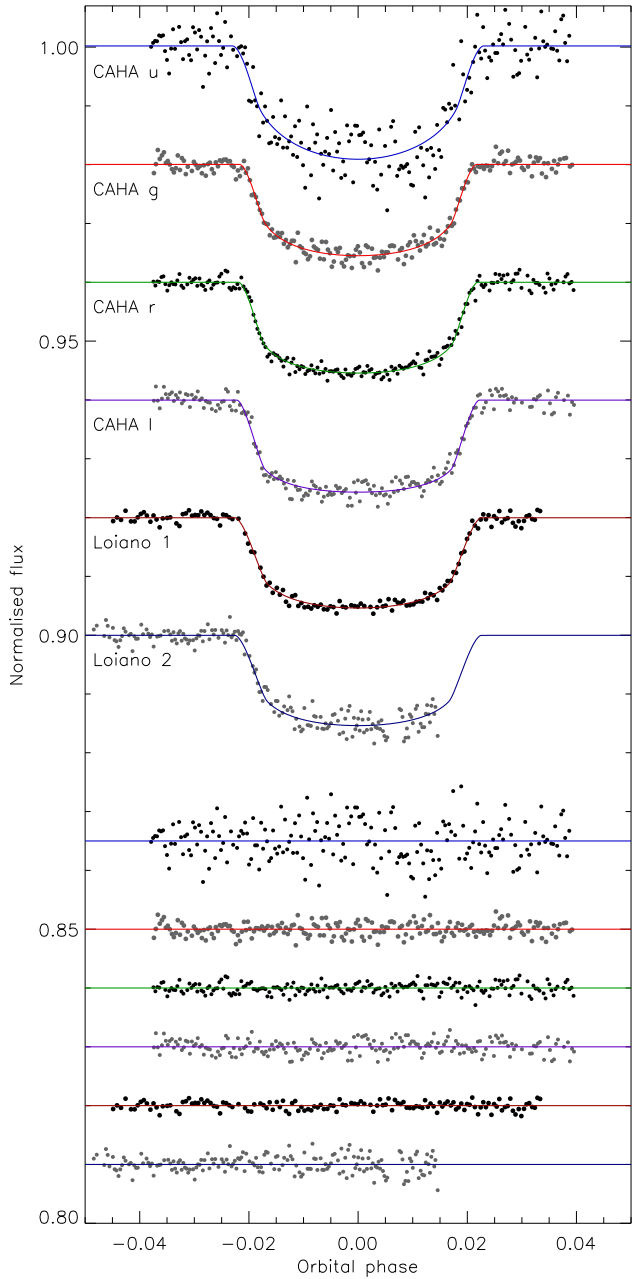}
\caption{\label{fig:lcfit} Phased light curves of HAT-P-5 compared to the best
{\sc jktebop} fits using the quadratic limb darkening law. The sources
of the light curves are labelled on the plot. The residuals of the fits are
plotted at the base of the figure, offset from zero.} \end{figure}

\begin{figure} \includegraphics[width=0.48\textwidth,angle=0]{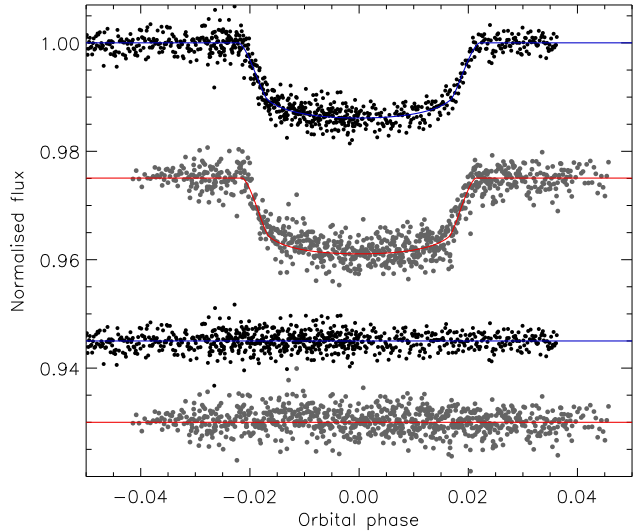}
\caption{\label{fig:lcfit2} As for Fig.\,\ref{fig:lcfit} but for the $z$ and
$R$ light curves of HAT-P-5 presented by \citet{Bakos+07apj2}.} \end{figure}

As a first step in our analysis we fitted all of the light curves individually using {\sc jktebop} (see below). The absolute values of the observational errors from our pipeline (which stem ultimately from the {\sc aper} subroutine) are optimistic, so were rescaled for each dataset to give a reduced $\chi^2$ of $\chir = 1$. The times of mid-transit were then measured, and their uncertainties estimated from Monte Carlo simulations. The second Loiano transit was discounted from this as the data cover only part of the transit event \citep[see][]{Gibson+09apj}.

To our times of minimum we added 14 timings obtained by amateur astronomers and made available on the AXA\footnote{Amateur Exoplanet Archive, {\tt http://brucegary.net/AXA/x.htm}} and TRESCA\footnote{The TRansiting ExoplanetS and CAndidates (TRESCA) website can be found at, {\tt http://var2.astro.cz/EN/tresca/index.php}} websites. We included only timings based on data of low scatter and covering a full transit. All timings whose timescale is not stated were assumed to be on the UTC system and converted to TDB.

The resulting measurements of transit midpoints were fitted with a straight line to obtain an orbital ephemeris. Cycle zero is specified to be the midpoint of the transit observed at CAHA. We found:
$$ T_0 = {\rm BJD(TDB)} \,\, 2\,455\,432.45510 (10) \, + \, 2.78847360 (52) \times E $$
where $E$ is the number of orbital cycles after the reference epoch and quantities in brackets denote the uncertainty in the final digit of the preceding number. The quality of the fit is $\chi^2_\nu = 1.46$, which implies that a linear ephemeris is not a perfect fit to the transit timings. There is no systematic deviation from the predicted transit times, so we do not regard this as sufficient evidence for transit timing variations. The \chir\ is instead interpreted as an indication that the errorbars on the timings are too small, which we have compensated for by increasing the uncertainties in the ephemeris above.

\subsection{Light curve modelling}                                                                                                  \label{sec:lc:lc}

\begin{table*} \caption{\label{tab:lcfit} Parameters of the {\sc jktebop} fits to the light
curves of HAT-P-5. The final parameters are the weighted means of the results for the six
datasets. Results from the literature are included at the base of the table for comparison.}
\begin{tabular}{l r@{\,$\pm$\,}l r@{\,$\pm$\,}l r@{\,$\pm$\,}l r@{\,$\pm$\,}l r@{\,$\pm$\,}l}
\hline
Source       & \mc{$r_{\rm A}+r_{\rm b}$} & \mc{$k$} & \mc{$i$ ($^\circ$)} & \mc{$r_{\rm A}$} & \mc{$r_{\rm b}$} \\
\hline
CAHA $g$-band           & 0.1407 & 0.0080 & 0.1130 & 0.0025 & 87.8  & 1.7  & 0.1264 & 0.0069 & 0.0143 & 0.0011 \\
CAHA $r$-band           & 0.1442 & 0.0077 & 0.1165 & 0.0022 & 87.7  & 1.8  & 0.1292 & 0.0066 & 0.0150 & 0.0010 \\
CAHA $I$-band           & 0.1404 & 0.0059 & 0.1134 & 0.0017 & 88.2  & 1.5  & 0.1262 & 0.0051 & 0.0143 & 0.0008 \\
Loiano $r$-band         & 0.1562 & 0.0084 & 0.1159 & 0.0026 & 86.1  & 1.1  & 0.1400 & 0.0073 & 0.0162 & 0.0011 \\
Bakos $z$-band          & 0.1464 & 0.0077 & 0.1094 & 0.0020 & 87.2  & 1.4  & 0.1320 & 0.0067 & 0.0144 & 0.0010 \\
Bakos $R$-band          & 0.1413 & 0.0095 & 0.1111 & 0.0025 & 87.4  & 1.9  & 0.1272 & 0.0084 & 0.0141 & 0.0012 \\
\hline
Final results & \mc{ } & \mc{ } & {\bf 87.18} & {\bf 0.61} & {\bf 0.1296} & {\bf 0.0027} & {\bf 0.01467} & {\bf 0.00041} \\
\hline
\citet{Bakos+07apj2}    &   \mc{0.1477}   & 0.1106 & 0.0006 & 86.75 & 0.44 & 0.133  & 0.003  &   \mc{0.01471}  \\
\citet{Torres++08apj}   &   \mc{0.1478}   & 0.1106 & 0.0006 & 86.75 & 0.44 & 0.1333 & 0.0033 &   \mc{0.01472}  \\
\hline \end{tabular} \end{table*}

The light curves were modelled using the {\sc jktebop}\footnote{{\sc jktebop} is written in {\sc fortran77} and the source code is available at {\tt http://www.astro.keele.ac.uk/$\sim$jkt/}} code as described in \citet{Me08mn}. The main information to be gleaned was the fractional radii of the star and planet, $r_{\rm A} = \frac{R_{\rm A}}{a}$ and $r_{\rm b} = \frac{R_{\rm b}}{a}$, where $a$ is the orbital semimajor axis and $R_{\rm A}$ and $R_{\rm b}$ are the true radii of the two objects. The fitted parameters were the sum and ratio of the fractional radii, $r_{\rm A}${$+$}$r_{\rm b}$ and $k = \frac{r_{\rm b}}{r_{\rm A}}$, and the orbital inclination, $i$. A mass ratio of 0.001 was adopted to govern the shapes of the biaxial ellipsoids representing the two component objects. Limb darkening (LD) was imposed using each of five laws and with three choices of whether the LD coefficients were fixed or fitted.
Uncertainties were calculated using both Monte Carlo simulations \citep{Me++04mn2} and a residual-permutation algorithm (as implemented by \citealt{Me08mn}).  The larger of the two values was retained for each output quantity.
The orbital eccentricity was fixed to zero \citep{Bakos+07apj2} and contaminating light was taken to be negligible \citep{Daemgen+09aa}.

The CAHA $g$, $r$ and $I$ data were solved individually, and the $u$-band light curve was ignored at this point due to its much greater scatter. The Loiano transits were solved simultaneously but with the orbital period fixed at the value determined in Section\,\ref{sec:lc:porb}. We also considered the $z$-band and $R$-band follow-up data presented in \citet{Bakos+07apj2}. In all six cases we found that the best solutions were obtained when fitting for the linear LD coefficient and fixing the nonlinear coefficient at a theoretically-predicted value (but perturbing it by $\pm$0.1 on a flat distribution during the Monte Carlo and residual-permutation simulations).

An overall set of photometric parameters was calculated for each dataset and is given in Table\,\ref{tab:lcfit}. Tables of individual fits for each light curve are given in the Supplementary Information which accompanies this work. The corresponding best fits are shown in Fig.\,\ref{fig:lcfit} for the data presented in this work and in Fig.\,\ref{fig:lcfit2} for the data from \citet{Bakos+07apj2}. The final photometric parameters are the weighted mean of the values for each dataset. At first glance the Loiano results are outliers, but this impression is not backed up by statistics: the \chir\ value of the agreement of the individual parameters with respect to the weighted mean is 1.5 for $k$ but smaller than 0.6 for the other photometric parameters. Table\,\ref{tab:lcfit} also shows a comparison with the results from \citet{Bakos+07apj2} and \citet{Torres++08apj}. These authors agree well with our final parameter values, but find errorbars which are notably smaller than ours for those datasets in common.

\subsection{Variation of planetary radius with wavelength}                                                                        \label{sec:lc:wave}

Once the final photometric parameters had been established, each light curve was fitted with $i$ and $r_{\rm A}$ fixed in order to investigate the possibility of a variation of $r_{\rm b}$ with wavelength. The resulting $r_{\rm b}$ values should have reliable relative uncertainties from which the common sources of error have been removed. We inflated the errorbars to account for systematic noise following the `$\beta$' approach and with groups of up to ten consecutive datapoints \citep[e.g.][]{Winn+07aj2}. One issue to bear in mind when fixing $i$ and $r_{\rm A}$ is that any variations in these parameters, or effects arising from systematic noise, will be concentrated in the remaining free parameters, including $r_{\rm b}$.

Whilst $r_{\rm b}$ {\em might} vary detectably with wavelength, the LD characteristics of the parent star certainly do. In order to make any errors in the theoretical representation of LD by stellar model atmospheres clear, we calculated $r_{\rm b}$ values with the linear LD coefficient included as a fitted parameter as well as solutions in which both coefficient were fixed to theoretical values. An excellent discussion of the treatment of LD in transit light curves is given by \citet{Howarth11}.

\begin{figure} \includegraphics[width=0.48\textwidth,angle=0]{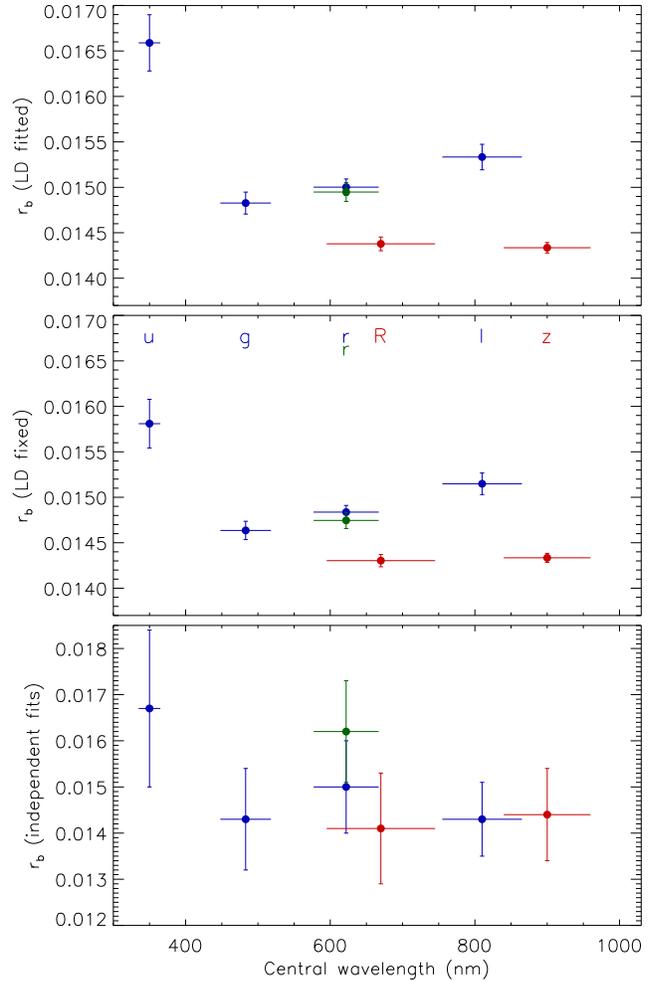}
\caption{\label{fig:radius} Variation of the fractional planetary radius
($r_{\rm b} = \frac{R_{\rm b}}{a}$) with wavelength. Results from the CAHA data
are shown in blue, from Loiano in green, and from the \citet{Bakos+07apj2} data
in red. The vertical lines represent the relative errors in the measurements and
the horizontal lines show the full widths at half maximum transmission of the
passbands used. \reff{The top panel gives the results for limb darkening fitted,
the middle panel for LD fixed at theoretically predicted values, and the lower
panel shows the results when the system parameters are not forced to be the same
for different light curves.} The filter designations are labelled at the top of
the lower panel.} \end{figure}

Fig.\,\ref{fig:radius} shows the variation of $r_{\rm b}$ with wavelength with the linear LD coefficient fitted (\reff{top} panel) and fixed (\reff{middle} panel). The corresponding results are given in Table\,\ref{tab:rbfit}. Clear variations with wavelength are seen, and are robust against the treatment of LD. In all cases fixing LD coefficients leads to $r_{\rm b}$ values which are smaller, an effect which is most pronounced for the $u$-band data. Changing the LD coefficients to their physical limits (i.e.\ the total LD at the limb of the star is zero or unity) cannot shift the $u$-band $r_{\rm b}$ into alignment with the other results. 

\reff{The referee raised concerns that correlated noise in the photometry could be influencing the observed variations in $r_{\rm b}$, and that by fixing $i$ and $r_{\rm A}$ to common values any systematic errors would be concentrated in $r_{\rm b}$. We have therefore shown in Fig.\,\ref{fig:radius} (bottom panel) the values of $r_{\rm b}$ obtained when $i$ and $r_{\rm A}$ are not constrained to be the same for each light curve. The differences between the three panels in Fig.\,\ref{fig:radius} therefore highlight the effect of adopting a common set of system parameters.}

So is the variation of $r_{\rm b}$ with wavelength real? The two $r$-band datasets are in good agreement, which is encouraging. The $u$-band result is very discrepant with those from the other bands, but the reality of this phenomenon is arguable. Of all optical wavelengths, photometry blueward of the Balmer jump suffers the most from the strength and variability of Earth-atmospheric extinction, although the altitude of CAHA (2168\,m above sea level) will mitigate that to some extent. $u$- and $U$-band photometry is also often handicapped by a lack of comparison stars, which is true in the current case (only one good comparison star compared to at least two for all other bands). Our CAHA observations were also taken at full moon (illuminated fraction 0.994) so the sky background was comparatively high in $u$ and $g$. Fig.\,\ref{fig:lcfit} shows that our $u$-band data have some correlated noise visible in the residuals.

A second suggestion of correlated noise affecting the $r_{\rm b}$ values is that those for the $z$- and $R$-band data are significantly below the values found from our new data. Whilst this could be interpreted as indicative of temporal variation in the properties of the system, it is most likely to herald the presence of systematic errors in the photometry. Further observations are needed to assess the significance of the $r_{\rm b}$ variations we see. For the remainder of this work we proceed under the assumption that the variations of $r_{\rm b}$ with wavelength are unconfirmed.

\subsubsection{Possible explanations of the radius variation}

If the observed variations of radius with wavelength were true, what would they imply? The $r_{\rm B}$ value from the $u$-band data is substantially larger than those in the other passbands, which is in the opposite direction to the predictions of \citet{Fortney+08apj}. Rayleigh scattering causes a larger radius at bluer wavelengths, which would fit the $u$ result but not the more modest trend seen in the $g$, $r$ and $I$ results. Similarly, the sulphur chemistry advocated by \citet{Zahnle+09apj} would lead to a larger radius in both $u$ and $g$ due to opacity of HS.

Next, one can calculate the size of the $u$-band radius difference in units of the atmospheric pressure scale height, $H$. The $r_{\rm b}$ at $u$ is 7\% (for fixed LD coefficients) or 11\% (linear LD coefficient fitted) larger than that for the $r$ band. This equates to a difference of 6100\,km or 9700\,km. For HAT-P-5\,b $H \approx 500$\,km (using data from \citealt{DepaterLissauer01book}). The radius variation is therefore roughly $12H$ or $19H$, which is large but not excessive. \citet[][their fig.\,14]{Sing+11mn} measured radii for HD\,189733\,b which approached $6H$ larger at 330\,nm than that measured by \citet{Pont+08mn} in the 1000--1025\,nm passband. Rayleigh scattering is therefore a plausible expanation.

Two further possibilities are apparent. Firstly, an unusally large transit depth was found for WASP-12 by \citet{Fossati+10apj} and interpreted as evidence for a extended exosphere surrounding the planet, with the possible presence of a disc formed from excreted material. The observations have subsequently been analysed in the context of a bow shock arising from the motion of the planet through the stellar wind \citep{Vidotto++10apj,Llama+11mn}. The observable consequences of this situation are that UV transits are deeper and begin earlier than those at redder wavelengths. A bow shock would be much weaker due to the larger orbital separation in the HAT-P-5 system ($0.04079 \pm 0.00080$\,AU; see below) compared to WASP-12 ($0.02293 \pm 0.00078$\,AU; \citealt{Hebb+09apj}; \citealt{Chan+11aj}). Whilst our observations satisfy the criterion of a deeper transit at bluer wavelengths, they do not indicate that the $u$ transit occurs earlier than the $g$, $r$ and $I$ ones (see Table\,\ref{tab:minima}).

\begin{table} \centering \caption{\label{tab:rbfit} Fractional planetary
radius values ($r_{\rm b} = \frac{R_{\rm b}}{a}$) found from the available
data with the linear limb darkening (LD) coefficient either fitted or fixed,
when the other photometric parameters are fixed at their known/final values.}
\begin{tabular}{l r@{\,$\pm$\,}l r@{\,$\pm$\,}l}
\hline
Source & \mc{$r_{\rm b}$ (LD fitted)} & \mc{$r_{\rm b}$ (LD fixed)} \\
\hline
CAHA $u$-band     & 0.01659 & 0.00031 & 0.01581 & 0.00027 \\
CAHA $g$-band     & 0.01483 & 0.00012 & 0.01464 & 0.00011 \\
CAHA $r$-band     & 0.01500 & 0.00009 & 0.01484 & 0.00007 \\
CAHA $I$-band     & 0.01533 & 0.00014 & 0.01515 & 0.00012 \\
Loiano $r$-band   & 0.01495 & 0.00010 & 0.01475 & 0.00009 \\
Bakos $z$-band    & 0.01434 & 0.00006 & 0.01433 & 0.00005 \\
Bakos $R$-band    & 0.01438 & 0.00008 & 0.01430 & 0.00007 \\
\hline \end{tabular} \end{table}

The second possibility is that the stellar surface exhibited starspots during our observations, which are not directly detectable becasue they were not occulted by the planet. This would act to increase the depth of the transit. If the temperature of the spot(s) is close enough to the \Teff\ of the star that they contribute significant flux at redder wavelengths, this would induce a wavelength dependence such that transits in the blue are deeper than those in the red. Such a phenomenon would explain both a deeper $u$-band transit and a greater transit depth overall compared to those in \citet{Bakos+07apj2}. However, it does not agree well with the observed trend in the $g$, $r$ and $I$ bands. The level of starspot activity would also have to be similar at both of our $r$-band epochs, which are separated by nine months, in order to explain their mutual agreement.

\begin{table*} \centering \caption{\label{tab:hatp5:model} Derived physical
properties of the HAT-P-5 system. For each set of physical properties the
following quantities are the same: $g_{\rm b} = 16.7 \pm 1.9$\mss,
$\rho_{\rm A} = 0.791 \pm 0.050$\psun\ and $\Teq = 1517 \pm 29$\,K.}
\begin{tabular}{l r@{\,$\pm$\,}l r@{\,$\pm$\,}l r@{\,$\pm$\,}l r@{\,$\pm$\,}l r@{\,$\pm$\,}l r@{\,$\pm$\,}l}
\hline
\ & \mc{This work} & \mc{This work} & \mc{This work} & \mc{This work} & \mc{This work} & \mc{This work} \\
\ & \mc{(dEB constraint)} & \mc{({\sf Claret} models)} & \mc{({\sf Y$^2$} models)} & \mc{({\sf Teramo} models)} & \mc{({\sf VRSS} models)} & \mc{({\sf DSEP} models)} \\
\hline
$K_{\rm b}$     (\kms) & 162.2   &   4.9    & 159.8   &   1.9    & 159.1   &   2.3    & 158.6   &   2.9    & 158.0   &   2.4    & 158.5   &   3.0    \\
$M_{\rm A}$    (\Msun) & 1.240   & 0.113    & 1.185   & 0.042    & 1.170   & 0.051    & 1.159   & 0.063    & 1.147   & 0.052    & 1.156   & 0.065    \\
$R_{\rm A}$    (\Rsun) & 1.161   & 0.044    & 1.144   & 0.029    & 1.139   & 0.029    & 1.135   & 0.031    & 1.132   & 0.029    & 1.135   & 0.032    \\
$\log g_{\rm A}$ (cgs) & 4.402   & 0.021    & 4.395   & 0.019    & 4.393   & 0.019    & 4.392   & 0.020    & 4.390   & 0.020    & 4.392   & 0.020    \\[2pt]
$M_{\rm b}$    (\Mjup) & 1.10    & 0.13     & 1.07    & 0.11     & 1.06    & 0.11     & 1.06    & 0.11     & 1.05    & 0.11     & 1.05    & 0.11     \\
$R_{\rm b}$    (\Rjup) & 1.279   & 0.053    & 1.260   & 0.039    & 1.254   & 0.040    & 1.250   & 0.042    & 1.246   & 0.040    & 1.250   & 0.042    \\
$\rho_{\rm b}$ (\pjup) & 0.494   & 0.067    & 0.501   & 0.066    & 0.503   & 0.067    & 0.505   & 0.067    & 0.507   & 0.067    & 0.505   & 0.067    \\
\safronov\             & 0.0580  & 0.0064   & 0.0589  & 0.0063   & 0.0591  & 0.0063   & 0.0593  & 0.0063   & 0.0595  & 0.0063   & 0.0594  & 0.0064   \\[2pt]
$a$               (AU) & 0.04166 & 0.00126  & 0.04104 & 0.00049  & 0.04086 & 0.00059  & 0.04073 & 0.00074  & 0.04059 & 0.00061  & 0.04070 & 0.00076  \\
\hline \end{tabular} \end{table*}

\begin{table*} \centering \caption{\label{tab:hatp5:final} Final physical properties of the
HAT-P-5 system, compared with results from the literature. Where two errorbars are given,
the first refers to the statistical uncertainties and the second to the systematic errors.}
\begin{tabular}{l r@{\,$\pm$\,}c@{\,$\pm$\,}l r@{\,$\pm$\,}l r@{\,$\pm$\,}l}
\hline
\ & \mcc{This work (final)} & \mc{\citet{Bakos+07apj2}} & \mc{\citet{Torres++08apj}} \\
\hline
$M_{\rm A}$    (\Msun) & 1.163    & 0.065    & 0.022     &   1.160 & 0.062   & \erc{1.157}{0.043}{0.081}       \\
$R_{\rm A}$    (\Rsun) & 1.137    & 0.032    & 0.007     &   1.167 & 0.049   & \erc{1.165}{0.046}{0.052}       \\
$\log g_{\rm A}$ (cgs) & 4.392    & 0.020    & 0.003     &   4.368 & 0.028   & \erc{4.368}{0.025}{0.031}       \\
$\rho_{\rm A}$ (\psun) & \mcc{$0.791 \pm 0.050$}         &      \mc{ }       & \erc{0.729}{0.058}{0.054}       \\[2pt]
$M_{\rm b}$    (\Mjup) & 1.06     & 0.11     & 0.01      &    1.06 & 0.11    & \erc{1.06}{0.11}{0.11}          \\
$R_{\rm b}$    (\Rjup) & 1.252    & 0.042    & 0.008     &   1.257 & 0.053   & \erc{1.254}{0.051}{0.056}       \\
$g_{\rm b}$     (\mss) & \mcc{$16.7 \pm  1.9$}           &    16.5 & 1.9     & \erc{16.6}{1.9}{1.8}            \\
$\rho_{\rm b}$ (\pjup) & 0.504    & 0.067    & 0.003     &    0.50 & 0.08    & \erc{0.50}{0.09}{0.08}          \\[2pt]
\Teq\              (K) & \mcc{$1517 \pm   29$}           &      \mc{ }       & \erc{1539}{33}{32}              \\
\safronov\             & 0.0592   & 0.0064   & 0.0004    &      \mc{ }       & \erc{0.0591}{0.0064}{0.0062}    \\
$a$               (AU) & 0.04079  & 0.00076  & 0.00026   & 0.04075 & 0.00076 & \erc{0.04071}{0.00049}{0.00097} \\
Age              (Gyr) & \ermcc{1.7}{3.5}{1.4}{0.4}{0.6} &     2.6 & 1.8     & \erc{2.6}{2.1}{1.4}             \\
\hline \end{tabular} \end{table*}

The spot activity of HAT-P-5\,A can be investigated using time-series photometry extending over a significant time interval. The SuperWASP archive \citep{Pollacco+06pasp,Butters+10aa} lists 48\,392 observations of HAT-P-5 over the years 2004 to 2010. We searched these data for periodicities indicative of spot-induced rotational modulation, using the method of \citet{Maxted+11pasp}. The periodograms of the data show only peaks near 1\,d and 28\,d, which we attribute to the diurnal and lunar cycles. We set a 95\%-confidence upper limit of 0.5\,mmag on rotational modulations, which confirms that HAT-P-5 is a quiet star.

In conclusion, we observe a different transit depth in $u$ compared to $g$, $r$ and $I$. We also find lower transit depths for the $R$ and $z$ observations obtained by \citet{Bakos+07apj2}. We have attempted to interpret these effects in the context of a variation of opacity with wavelength in the planetary atmosphere, the presence of a bow shock, or spot activity on the host star. None of these possibilities provide a completely satisfactory description of the data, but Rayleigh scattering remains plausible. Our preferred explanation is that the differences in transit depth are due to systematic effects in the available photometry.

\subsection{Physical properties of the HAT-P-5 system}                                                                         \label{sec:properties}

We determined the physical properties of the HAT-P-5 system using the approach described in \citet{Me09mn}. Starting with the measured photometric parameters and known orbital velocity amplitude of the star ($K_{\rm A} = 138 \pm 14$\ms; \citealt{Bakos+07apj2}), we guessed a velocity amplitude for the planet ($K_{\rm b}$) and used standard formulae \citep[e.g.][]{Hilditch01book} to obtain an estimate of the properties of the system. These comprise the mass, radius, surface gravity and mean density for the star ($M_{\rm A}$, $R_{\rm A}$, $\log g_{\rm A}$ and $\rho_{\rm A}$) and planet ($M_{\rm b}$, $R_{\rm b}$, $g_{\rm b}$ and $\rho_{\rm b}$), and the orbital semimajor axis ($a$). We then interpolated within the predictions of theoretical stellar models to find the effective temperature (\Teff) and $R_{\rm A}$ corresponding to the estimated $M_{\rm A}$ and measured metallicity ($\FeH = 0.24 \pm 0.15$; \citealt{Bakos+07apj2}). The value of $K_{\rm b}$ was then iteratively refined to maximise the agreement between the estimated and observed \Teff\ ($5960 \pm 100$\,K; \citealt{Bakos+07apj2}), and the estimated $\frac{R_{\rm A}}{a}$ and known $r_{\rm A}$. This process was performed for a grid of ages starting from the zero-age main sequence and stepping in 0.01\,Gyr chunks until the star had evolved to a $\logg$ less than $3.5$. The final result was a set of physical properties and a system age which gave the best agreement with the known \Teff\ and $r_{\rm A}$.

The random errors in this process were propagated via a perturbation analysis \citep{Me++05aa}. Our use of theoretical models inevitably incurs a systematic error due to the dependence on stellar theory. We assessed these systematic errors from the interagreement of sets of physical properties calculated using each of five independent sets of stellar models \citep{Me11mn}. Our final result for each physical property corresponds to the unweighted mean of the five different values, whilst its statistical error is the largest of the five possibilities and its systematic error is the standard deviation of the values from the five models. Note that three quantities are independent of stellar theory so are not troubled by systematic errors: $g_{\rm b}$ \citep{Me++07mn}, $\rho_{\rm A}$ \citep{SeagerMallen03apj} and \Teq\ \citep{Me09mn}.

It is possible to avoid using stellar models by recourse to calibrations based on empirical measurements of stars in eclipsing binary systems \citep{Me09mn}. We adopted the approach outlined by \citet{Enoch+10aa} but with the improved calibration coefficients calculated by \citet{Me11mn}. The set of physical properties resulting from this method are shown in Table\,\ref{tab:hatp5:model}, along with the five sets arriving from the use of stellar models. Our final properties are given in Table\,\ref{tab:hatp5:final}. The results obtained by \citet{Bakos+07apj2} and \citet{Torres++08apj} are in very close agreement with our own, despite being based on much sparser photometric data.


\section{Summary}                                                                                                                 \label{sec:summary}

We present photometry of four transit events in the HAT-P-5 extrasolar planetary system, obtained using telescope-defocussing techniques and reaching scatters as low as 0.75\,mmag per point. One of these transits was observed in four passbands simultaneously, using the BUSCA imager on the CAHA 2.2\,m telescope. We used these data to improve the measured orbital ephemerides and physical properties of the system. HAT-P-5 is well-characterised, although it would benefit from further observations to refine the spectroscopic orbit and atmospheric properties of the host star. The planet is slightly too large to match the predicted radii of coreless gaseous planets \citep{Fortney++07apj,Baraffe++08aa}, confirming its status as an inflated hot Jupiter. HAT-P-5 becomes the sixtieth TEP system to be studied as part of the {\it Homgeneous Studies} project \citep{Me08mn} and its physical properties have been placed in the Transiting Extrasolar Planets Catalogue\footnote{TEPCat: {\tt http://www.astro.keele.ac.uk/$\sim$jkt/tepcat/}}

Both components of HAT-P-5 have similar temperatures to those in the prototype transiting system HD\,209458, so the planets are expected to have similar atmospheric characteristics. We used our multi-band photometry to search for variations of the measured planetary radius with wavelength, which could indicate the presence of Rayleigh scattering or the opacity of specific molecules. We find that the radius in the $u$ band is significantly larger than at the other optical wavelengths, by either 12 or 19 pressure scale heights depending on the treatment of limb darkening when fitting the light curves. We also find that the planet radius measured in earlier $R$ and $z$-band data is smaller than that from our data. These phenomena are most likely due to the presence of systematic errors in the photometric data. Alternative possibilities include Rayleigh scattering and temporal changes in the system properties, but explanations involving starspots or a bow shock do not match existing observations.


\section*{Acknowledgments}

Based on observations collected at the Centro Astron\'omico Hispano Alem\'an (CAHA) at Calar Alto, Spain, operated jointly by the Max-Planck Institut f\"ur Astronomie and the Instituto de Astrof\'{\i}sica de Andalucía (CSIC), and on observations obtained with the 1.5\,m Cassini telescope at Loiano Observatory, Italy. The reduced light curves presented in this work will be made available at the CDS ({\tt http://cdsweb.u-strasbg.fr/}) and at {\tt http://www.astro.keele.ac.uk/$\sim$jkt/}. JS acknowledges financial support from STFC in the form of an Advanced Fellowship. We thank Roberto Gualandi for his technical assistance at the Cassini telescope, Simona Ciceri for taking part to these observations, and Nikolay Nikolov for suggesting the bow-shock explanation. The following internet-based resources were used in research for this paper: the ESO Digitized Sky Survey; the NASA Astrophysics Data System; the SIMBAD database operated at CDS, Strasbourg, France; and the ar$\chi$iv scientific paper preprint service operated by Cornell University.

\bibliographystyle{mn_new}

\end{document}